\documentclass[10pt]{article} 

\usepackage{fullpage}
\usepackage{array}
\usepackage[dvips]{graphicx}
\usepackage{epic, eepic}
\usepackage{amssymb,amsmath}
\usepackage{latexsym}
\usepackage{subfigure}
\usepackage[usenames]{color}
\usepackage{multirow}
\usepackage{gastex}
\usepackage{subfigure}
\usepackage{ctable}
\usepackage{mdwlist}

\newcounter{compressEnum}

\newtheorem{theo}{Theorem} 
\newenvironment{theorem}{\begin{theo}}{\end{theo}}

\newcommand{\always}{\textrm{always}}
\newcommand{\nxt}{\textrm{next}}
\newcommand{\eventually}{\textrm{eventually}}
\newcommand{\untl}{\textrm{until\_}}
\newcommand{\bfore}{\textrm{before}}
\newcommand{\hbusreq}{\textrm{HBUSREQ}}
\newcommand{\busreq}{\textrm{BUSREQ}}
\newcommand{\hbusreqi}{\textrm{HBUSREQ}$_{\textrm{i}}$}
\newcommand{\hlock}{\textrm{HLOCK}}
\newcommand{\hlocki}{\textrm{HLOCK}$_{\textrm{i}}$}
\newcommand{\hready}{\textrm{HREADY}}
\newcommand{\hgrant}{\textrm{HGRANT}}
\newcommand{\hgrantz}{\textrm{HGRANT0}}
\newcommand{\hgranti}{\textrm{HGRANT}$_{\textrm{i}}$}
\newcommand{\hmastlock}{\textrm{HMASTLOCK}}
\newcommand{\hmaster}{\textrm{HMASTER}}
\newcommand{\haddr}{\textrm{HADDR}}
\newcommand{\haddri}{\textrm{HADDR}$_{\textrm{i}}$}
\newcommand{\htrans}{\textrm{HTRANS}}
\newcommand{\hburst}{\textrm{HBURST}}
\newcommand{\hresp}{\textrm{HRESP}}
\newcommand{\hwrite}{\textrm{HWRITE}}
\newcommand{\hwdata}{\textrm{HWDATA}}
\newcommand{\hwdatai}{\textrm{HWDATA}$_{\textrm{i}}$}
\newcommand{\hrdata}{\textrm{HRDATA}} 
\newcommand{\hrdatai}{\textrm{HRDATA}$_{\textrm{i}}$}
\newcommand{\hsze}{\textrm{HSIZE}}
\newcommand{\hsel}{\textrm{HSEL}}
\newcommand{\hsplit}{\textrm{HSPLIT}}
\newcommand{\granted}{\textrm{GRANTED}}
\newcommand{\decide}{\textrm{DECIDE}}
\newcommand{\start}{\textrm{START}}
\newcommand{\locked}{\textrm{LOCKED}}

\newcommand{\single}{\textrm{SINGLE}}
\newcommand{\incr}{\textrm{INCR}}
\newcommand{\incrf}{\textrm{INCR4}}
\newcommand{\nonseq}{\textrm{NONSEQ}}
\newcommand{\seq}{\textrm{SEQ}}
\newcommand{\word}{\textrm{WORD}}
\newcommand{\busy}{\textrm{BUSY}}
\newcommand{\idle}{\textrm{IDLE}}
\newcommand{\okay}{\textrm{OKAY}}
\newcommand{\splt}{\textrm{SPLIT}}
\newcommand{\error}{\textrm{ERROR}}
\newcommand{\retry}{\textrm{RETRY}}
\newcommand{\ful}{\textrm{FULL}}
\newcommand{\emty}{\textrm{EMPTY}}

\newcommand{\amba}{\textrm{AMBA}}
\newcommand{\ahb}{\textrm{AHB}}
\newcommand{\abc}{\textrm{ABC}}
\newcommand{\psl}{\textrm{PSL}}
\newcommand{\eqi}{\textrm{= i}}
\newcommand{\eqj}{\textrm{= j}}
\newcommand{\eqz}{\textrm{= 0}}
\newcommand{\low}{\textrm{LOW}}
\newcommand{\high}{\textrm{HIGH}}
\newcommand{\reqvld}{\textrm{REQ\_VLD}}
\newcommand{\wrt}{\textrm{WR}}
\newcommand{\re}{\textrm{RD}}
\newcommand{\leno}{\textrm{LEN1}}
\newcommand{\lenf}{\textrm{LEN4}}
\newcommand{\lenu}{\textrm{LENX}}
\newcommand{\inaddr}{\textrm{IN\_ADDR}}
\newcommand{\addr}{\textrm{ADDR}}
\newcommand{\di}{\textrm{DI}}
\newcommand{\dout}{\textrm{DO}}
\newcommand{\inaddri}{\textrm{IN\_ADDR}$_{\textrm{i}}$}
\newcommand{\indata}{\textrm{IN\_DATA}}
\newcommand{\indatai}{\textrm{IN\_DATA}$_{\textrm{i}}$}
\newcommand{\outdata}{\textrm{OUT\_DATA}}
\newcommand{\outdatai}{\textrm{OUT\_DATA}$_{\textrm{i}}$}
\newcommand{\last}{\textrm{LAST}}
\newcommand{\strt}{\textrm{START}}
\newcommand{\reqaddr}{\textrm{REQ\_ADDR}}
\newcommand{\reqwrdata}{\textrm{REQ\_WR\_DATA}}
\newcommand{\recrddata}{\textrm{REC\_RD\_DATA}}

\newtheorem{lemm}{Lemma}

\makeatletter

\begingroup \catcode `|=0 \catcode `[= 1
\catcode`]=2 \catcode `\{=12 \catcode `\}=12
\catcode`\\=12 |gdef|@xcomment#1\end{comment}[|end[comment]]
|endgroup

\def\@comment{\let\do\@makeother \dospecials\catcode`\^^M=10\def\par{}}

\def\begincomment{\@comment\@xcomment}

\makeatother

\usepackage{url}

\title{Synthesis of AMBA AHB from Formal Specification}
\author{Yashdeep Godhal \and Krishnendu Chatterjee \and Thomas A. Henzinger\medskip \\
\and IST Austria (Institute of Science and Technology Austria)
}

\date{}
\begin{document}
\pagestyle{plain}
\maketitle

\begin{abstract}
{\small
The standard procedure for hardware design consists of describing circuit 
in a hardware description language at logic level followed by extensive 
verification and logic-synthesis. 
However, this process consumes significant time and needs a lot of effort. 
An alternative is to use formal specification language as 
a high-level hardware description language and synthesize hardware from 
formal specification.
In~\cite{DBLP:conf/date/BloemGJPPW07} formal specifications for AMBA AHB 
Arbiter were presented and synthesized. 
Our contributions are as follows:
(1)~We present more complete and compact formal specifications for the
AMBA AHB Arbiter, and obtain significant (order of magnitude) improvement 
in synthesis results (both with respect to time and the number of gates of 
the synthesize circuit);
(2)~we present formal specification and synthesize to generate compact 
circuits for the remaining two components of the AMBA AHB protocol, namely, 
the AMBA AHB Master and AMBA AHB Slave;  and
(3)~from the lessons learnt we present few principles for writing formal 
specifications for efficient hardware synthesis.
Thus with systematically written complete formal specifications we are able to 
automatically synthesize an important and widely used industrial protocol.
}
\end{abstract}

\section{Introduction}

\noindent{\bf Hardware design flow.}
The first step in traditional standard industrial procedure of hardware design 
is the decription of a circuit in hardware description language. 
This step is followed by extensive verification and subsequently by logical 
synthesis. 
The outcome of logical syntehsis is gate level implementation of circuit. 
Among the above steps of design, verification and logical synthesis, the 
verification step is most time consuming process and requires a lot of effort.
An alternative approach is to automatically synthesize the circuit from 
formal specification.

\smallskip\noindent{\bf Synthesis from formal specification.}
Historically, automatic synthesis of digital designs from logical temporal 
specifications has been considered as one of the most challenging problems in 
circuit design. 
The problem was first presented by Church~\cite{Church63} and 
several methods have been proposed as solutions such as~\cite{1969} 
and in~\cite{540412}. 
The problem was considered again in~\cite{75293} in the context of 
synthesizing reactive modules from a specification given in Linear Temporal 
Logic (LTL). 
The method proposed in~\cite{75293} for a given LTL specification $\varphi$ 
starts by constructing a B\"uchi automaton which is converted into a 
deterministic Rabin automaton. This translation may require a doubly 
exponential complexity in the size of $\varphi$.
The high complexity established in~\cite{75293} caused synthesis to be deemed 
hopelessly intractable and discouraged practitioners from attempting to use 
it for system development. 
Yet, there exist several interesting cases where, if the specification of 
the design to be synthesized is restricted to simpler automata or 
partial fragments of LTL, it has been shown that the synthesis problem can 
be solved in polynomial time. 
Major progress has been achieved in~\cite{DBLP:conf/vmcai/PitermanPS06}, 
which shows that designs can be automatically synthesized from LTL formulas 
belonging to the class of generalized reactivity of rank 1 (GR(1)), 
in time $N^{3}$ where $N$ is the size of the state space of the design. 
The class GR(1) covers the vast majority of properties that appear in 
specifications of circuits. 
The approach of~\cite{DBLP:conf/vmcai/PitermanPS06} was implemented 
by~\cite{DBLP:conf/date/BloemGJPPW07} in a tool called 
Anzu~\cite{DBLP:conf/cav/JobstmannGWB07}. 
Anzu produces not only a BDD representing a set of possible implementations, 
but also an actual circuit.

\smallskip\noindent{\bf AMBA AHB Protocol.} 
In this work we study the automatic synthesis of an important and 
widely used industrial protocol, namely, \emph{AMBA AHB} protocol.
ARM's \emph{Advanced Microcontroller Bus Architecture} (AMBA)~\cite{arm} 
specification defines an on chip communications standard for designing 
high-performance embedded microcontrollers. AMBA is today the de-facto 
standard for embedded processors because it is well documented 
and can be used without royalties.  It is widely used in network 
interconnect chips, RAM and Flash memory controllers, DMA controllers, 
level2 cache controllers and SoCs including application processors used in
portable mobile devices like smartphones, and a few industrial examples 
of its use are IXP42X Product Line of Intel Network Processors, 
Infineon gateway controller ADM5120.
The most important bus defined within the \amba\ specification is 
\emph{Advanced High-performance Bus \ahb}. 
The \ahb\ acts as the high-performance system backbone bus. 
\ahb\ supports the efficient connection of processors, on-chip memories, 
DMA controllers and off-chip external memory interfaces.
The \amba\ \ahb\ design consists of the following main components:
(a)~AHB Arbiter; (b)~AHB Master and (c)~AHB Slave.
In this work we synthesize the above three components of the 
\amba\ \ahb\ protocol.

\smallskip\noindent{\bf Our contributions.} The contributions of this work 
are as follows:
\begin{enumerate}
\item In~\cite{DBLP:conf/date/BloemGJPPW07} 
and~\cite{DBLP:journals/entcs/BloemGJPPW07} the synthesis of only 
AMBA AHB Arbiter was studied. 
We present more complete and compact formal specifications for the
AMBA AHB Arbiter, and obtain significant (order of magnitude) improvement 
in synthesis results (both with respect to time and the number of gates of 
the synthesized circuit).
\item We present, for the first time, the formal specifications for the 
AMBA AHB Master and AMBA AHB Slave (the remaining two components of 
the protocol). We are able to synthesize very compact circuits from our 
formal specifications.
Thus we are able to completely synthesize an important and widely used 
industrial protocol by systematically writing the formal specifications.

\item From the lessons  that we have learnt in the process of rewriting 
specifications to obtain efficient synthesis result, we present few principles 
for writing formal specifications for efficient hardware synthesis.
\end{enumerate}

\section{Preliminaries}
In this section we present preliminaries related to specification language and 
synthesis.

\subsection{Property Specification Language} 
We will use \emph{Property Specification Language (PSL)} to express 
specifications (a detailed description of PSL can be found 
in~\cite{1209692}). 
The specifications presented in this paper are easy to follow for readers 
familiar with LTL. 
In particular, \always, \eventually, and \nxt\ correspond to $G$, $F$, and $X$, 
respectively. The \untl\ operator requires the first operand to hold either 
forever or up to and including the time that the second operand holds. 
The construct $(\Phi\  \bfore\ \Psi)$ is equivalent to 
$(\neg\Psi\  \untl\  \Phi)$.
We also use one operator that is not in PSL: $(\Phi\ \untl[i]\ \Psi)$ means 
that $\Phi$ holds either forever or up to and including the $i^{th}$ time that 
$\Psi$ holds. 

\subsection{Synthesis of GR(1) Properties} 
We briefly review the results presented in~\cite{DBLP:conf/vmcai/PitermanPS06} 
on synthesizing GR(1) properties. We are interested in the question of 
\emph{realizability} of PSL specifications (cf~\cite{DBLP:conf/popl/PnueliR89}).
Assume two sets of Boolean variables $X$ and $Y$. Intuitively $X$ is the set of
input variables controlled by the environment and $Y$ is the set of system 
variables. \emph{Realizability} amounts to checking whether there exists 
an \emph{open controller} that satisfies the specification. Such a controller 
can be represented as an automaton which, at any step, reads values of the $X$ 
variables and outputs values for the $Y$ variables.

Here we concentrate on a subset of PSL for which realizability and synthesis 
can be solved efficiently. The specifications we consider are of the form 
$\phi = \phi^{e} \to \phi^{s}$. We require that $\phi^{\alpha}$ for 
$\alpha \in \{e, s\}$ can be rewritten as a conjunction of the following parts.
\begin{itemize}
\item $\phi_{i}^{\alpha}$ - a Boolean formula which characterizes the initial 
states of the implementation.

\item $\phi_{t}^{\alpha}$ - a formula of the form $\wedge_{i} (\always\ B_{i})$ 
where each $B_{i}$ is a Boolean combination of variables from $X \cup Y$ and 
expressions of the form (\nxt\ $v$) where $v \in X$ if $\alpha = e$, and 
$v \in X \cup Y$ otherwise. 

\item $\phi_{g}^{\alpha}$ - has the form 
$\wedge_{i \in I}$ (\always\ \eventually\ $B_{i}$) where each $B_{i}$ is a 
Boolean formula.
\end{itemize}
In order to allow formulas of other forms 
(e.g. \always $(p \to (q$ \untl $ $ $r$)) where $p$, $q$, and $r$ are Boolean), 
we augment the set of variables by adding deterministic monitors. 
Deterministic monitors are variables whose behavior is deterministic according 
to the choice of the inputs and the outputs. These monitors follow the truth 
value of the expression nested inside the \always\ operator. 
We rewrite these types of formulas to the form (\always\ \eventually\  $b$) 
where $b$ is a Boolean formula using the variables of the monitor. 
It should be noted that even with these restrictions, all possible 
(finite state) designs can be expressed as a set of properties.

We reduce the realizability problem of a PSL formula to the decision of the 
winner in a two-player game played between system and environment. 
The goal of the system is to satisfy the specification regardless of the 
actions of the environment. 
A \emph{game structure} is a multigraph whose nodes are all the truth 
assignments to $X$ and $Y$. 
A node $v_1$ is connected by edges to all the nodes 
$v_2$ such that the truth assignments to $X$ and $Y$ satisfy 
$\phi_{t}^{e} \wedge \phi_{t}^{s}$ , where $v_1$ supplies the assignments 
to the current values and $v_2$ to the next values. 
We then group all the edges that agree on the assignment of $X$ in $v_2$ 
to one multi-edge. A play starts by 
the environment choosing an assignment to $X$ and the system choosing a state 
in $\phi_{i}^{e} \wedge \phi_{i}^{s}$ that agrees with this assignment. 
A play proceeds by the environment choosing a multi-edge and the system 
choosing one of the nodes connected to this multi-edge. 
The system wins if this interaction produces an infinite play that satisfies 
$\phi_{g}^{e} \to \phi_{g}^{s}$.

We \emph{solve} the game to decide whether the game is winning for the 
environment or the system. If the environment wins, then the specification 
is \emph{unrealizable}. If the system wins, then we \emph{synthesize} 
a winning strategy. This strategy, a BDD, is a nondeterministic representation 
of a working implementation. 
The following theorem summarizes the result of synthesis of PSL specifications.

\begin{theorem} \label{Theorem 1}~\cite{DBLP:conf/vmcai/PitermanPS06} 
Given sets of variables $X$ and $Y$ and a PSL formula $\phi$ of the form 
presented above with $m$ and $n$ conjuncts, we can determine using a 
symbolic algorithm whether $\phi$ is realizable in time proportional to 
$O(mn2^{d+|X |+|Y |})$, where $d$ is the number of variables added by the 
monitors for $\phi$. 
\end{theorem}

\subsection{Generating circuits from BDDs} 
We briefly review the results presented 
in~\cite{DBLP:journals/entcs/BloemGJPPW07} on generating circuits from BDDs. 
The strategy is a BDD over the variables $X$, $Y$, $X^{'}$ and $Y^{'}$, 
where $X$ are input variables, $Y$ are output variables and the primed 
versions represent next state variables. The corresponding circuit contains
$|X | + |Y|$ flipflops to store the values of the inputs and outputs in the 
last clock tick. 
In every step, the circuit reads the next input values $X^{'}$ and determines 
the next output values using combinational logic with inputs 
$I = X \cup Y \cup X^{'}$ and outputs $O = Y^{'}$. The strategy does not 
prescribe a unique combinational output for every combinational input. 
In most cases, multiple outputs are possible, in states that are not 
reachable (assuming that the system adheres to the strategy), no outputs 
may be allowed.

We write $o \in O$ for a combinational output and $i \in I$ for a 
combinational input. The strategy is denoted by $S$ and $O \setminus o$ is 
the set of combinational outputs excluding output $o$. For every combinational 
output $o$ we construct a function $f$ in terms of $I$ that is compatible with 
the given strategy BDD. The algorithm proceeds through the combinational 
outputs $o$ one by one: First, we build $S^{'}$ to get a BDD that restricts 
only $o$ in terms of $I$. Then we build the \textit{positive and negative 
cofactors} $(p,n)$ of $S^{'}$ with respect to $o$, that is, we find the sets 
of inputs for which $o$ can be 1 (0, respectively). For the inputs that occur 
in the positive and in the negative cofactor, both values are allowed. 
The combinational inputs that are neither in the positive nor in the negative 
cofactor are outside of the winning region and thus represent situations that 
cannot occur (as long as the environment satisfies the assumptions). 
Thus, $f$ has to be 1 in $p \wedge \neg n$ and 0 in $\neg p \wedge n $, 
which give us the set of care states. We minimize the positive cofactors 
with the care set to obtain the function $f$. 
Finally, we substitute variable $o$ in $S$ by $f$, and proceed with the next 
variable. 
The substitution is necessary since a combinational outputs may be related.

The resulting circuit is constructed by writing the BDDs for the functions 
using CUDD's DumpBlif command~\cite{cuddtool}. 
We then optimize the result using ABC~\cite{abctool} and map it to a library 
of standard cells. 
We also use ABC to estimate the number of gates needed.

\section{AMBA AHB Protocol}
In this section we describe the details of the main components of the 
\amba\ \ahb\ protocol.
ARM's \emph{Advanced Microcontroller Bus Architecture} (AMBA)~\cite{arm} 
specification defines an on chip communications standard for designing 
high-performance embedded microcontrollers. 
The most important bus defined within the \amba\ specification is 
\emph{Advanced High-performance Bus}. 
The \ahb\ acts as the high-performance system backbone bus. 
\ahb\ supports the efficient connection of processors, on-chip memories, 
DMA controllers and off-chip external memory interfaces.
The \amba\ \ahb\ design contains the following components:

\smallskip\noindent{\bf AHB master}: A bus master is able to initiate read 
and write operations by providing an address and control information. 
Only one bus master is allowed to actively use the bus at any one time. 

\smallskip\noindent{\bf AHB slave}: 
A bus slave responds to a read or write operation within a given 
address-space range. 
The bus slave signals back to the active master the success, failure or waiting
of the data transfer.

\smallskip\noindent{\bf AHB arbiter}: The bus arbiter ensures that only 
one bus master at a time is allowed to initiate data transfers. 
Even though the arbitration protocol is fixed, any arbitration algorithm, 
such as highest priority or fair access can be implemented depending on the 
application requirements.

\smallskip\noindent{\bf AHB decoder}: The AHB decoder is used to decode 
the address of each transfer and provide a select signal for the slave 
that is involved in the transfer.

Consider an AHB system with arbiter, masters and slaves. 
Every slave shall have some address range. 
AHB decoder receives address as input, checks in which range that address 
lies and provides select signal for slave that corresponds to this address. 
In essence, it works as a de-multiplexer. 
For a system with single slave, the select signal shall always be high, 
if valid address is put on bus. 
Hence we consider the synthesis of the main components of 
AHB design i.e. AHB Master, AHB Slave and AHB Arbiter.

\subsection{AHB Arbiter}
The role of the arbiter in an AMBA system is to control which master has 
access to the bus. 
Every bus master has a REQUEST/GRANT interface to the arbiter and the arbiter 
uses a prioritization scheme to decide which bus master is currently the 
highest priority master requesting the bus.
Each master also generates an HLOCK$_{\textrm{x}}$ signal which is used to 
indicate that the master requires exclusive access to the bus. 
The arbitration protocol is not specified and can be defined for each 
application. 

\subsection{AHB Master}
Function of AHB master is to do read and write operations. 
Before initiating any transfer, it sends a request to arbiter for accessing bus.
Once arbiter grants the bus, master initiates read/write operation by 
providing address and control information. 
Master 0 is the \emph{default master} and is selected whenever there are no 
requests for the bus.

\subsection{AHB Slave}
An AHB bus slave responds to transfers initiated by bus masters within the 
system. 
The slave uses a select signal HSEL$_{\textrm{x}}$ from the decoder to 
determine when it should respond to a bus transfer. 
All other signals required for the transfer, such as the address and control 
information, will be generated by the bus master. 

The AHB is a pipelined bus. This means that different masters can be in 
different stages of communication. At one instant, multiple masters can 
request the bus, while another master transfers address information, 
and a yet another master transfers data. 
A bus \emph{access} can be a \emph{single} transfer or a \emph{burst}, 
which consists of a specified or unspecified number of transfers. 
Access to the bus is controlled by the arbiter. 
All devices that are connected to the bus are Moore machines, that is, the 
reaction of a device to an action at time $t$ can only be seen by the other 
devices at time $t+1$.

\section{AMBA AHB Arbiter Synthesis}
In this section we present our results related to synthesis of AHB arbiter. 
We first present the arbiter signals, then present the formal specifications
and our result for synthesis.

\subsection{AHB Arbiter Signals}
\begin{figure}[h]
    	\includegraphics[width=\linewidth, height=0.4\textheight]{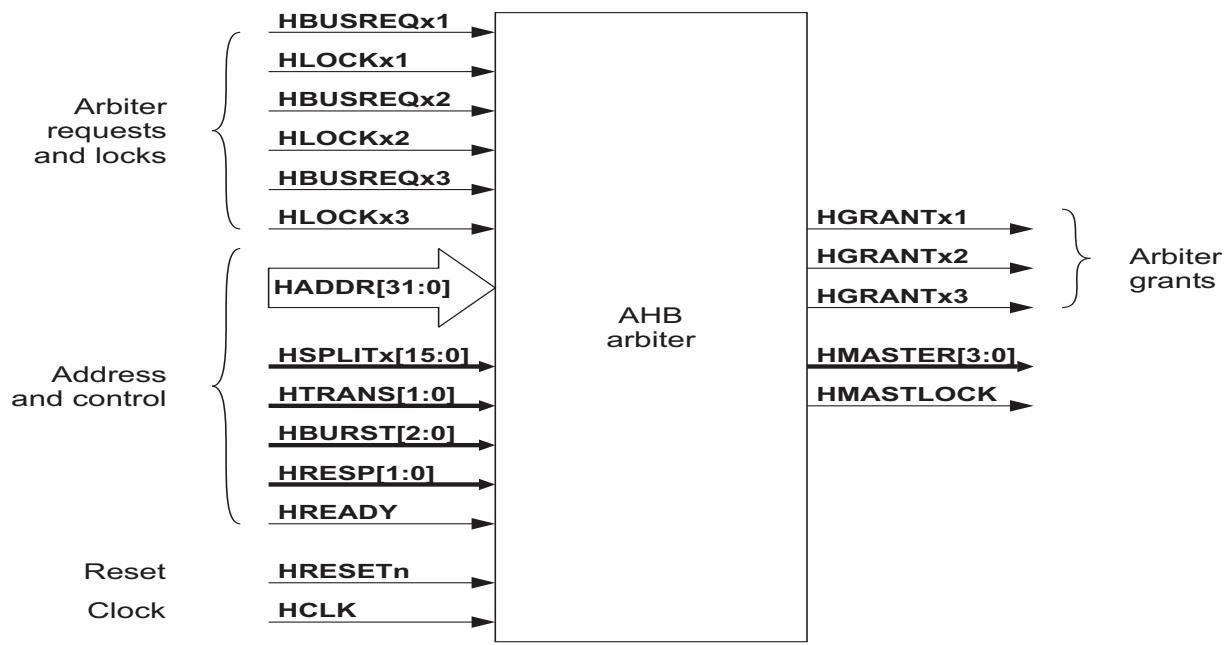}
     	\caption{\small AHB Arbiter~\cite{arm}}\label{fig1}
\end{figure}
Figure~\ref{fig1} shows AHB arbiter signals. 
The description of these signals are 
as follows (the notation S[n:0] denotes an (n+1)-bit signal):
\begin{itemize*}
\item \hbusreqi\ - A signal from bus master $i$ to the bus arbiter which 
indicates that the bus master requires access to the bus.
\item \hlocki\ - Indicates that the master requires locked access to the bus. 
No other master should be granted the bus until this signal is lowered.
\item \hready - This signal is driven by the bus slave. 
It indicates that a transfer has finished on the bus. 
This signal may be lowered to extend a transfer.
\item \hgranti\ - This signal indicates that if \hready\ is high, then \hmaster \eqi\ 
will hold in the next tick. 
\item \hmastlock\ - Indicates that the current master is performing a locked 
sequence of transfers.
\item \hmaster[3:0] - These signals from the arbiter indicate which bus master 
is currently performing a transfer.
\end{itemize*}
The following signals are multiplexed using \hmaster\ as the control signal. 
For example, although every master has an address bus, only the address 
provided by the currently active master is visible on \haddr.
\begin{itemize*}
\item \haddr[31:0] - These signals indicate the address where read or write 
transaction will take place.
\item \hburst[1:0] - One of \single\ (a single transfer), \incr\ (unspecified 
length burst) or \incrf\ (burst of four transfers). Though the standard allows 
for burst of eight and sixteen transfers too but we have not taken them into 
account. That would lengthen the specification.
\item \htrans[1:0] - Indicates the type of the current transfer, which can be 
\nonseq, \seq\ or \idle. The standard allows for \busy\ transfers also. \htrans\ = \busy\ 
indicates that master wants to introduce some delay during ongoing transfer. 
This is an optional feature. For simplicity we have left this feature out. 
\end{itemize*}
Furthermore, as an optional feature of the AHB, a slave is allowed to split a 
burst access and request that it be continued later (signals \hresp\ and \hsplit\ shown 
in Figure~\ref{fig1} serve that purpose). We have left this feature out for 
simplicity. 

Both optional features i.e. SPLIT and \busy\ transfers are also not considered
 in~\cite{DBLP:conf/date/BloemGJPPW07} while writing specifications for \ahb\ Arbiter.
Though they can be handled by this approach but that would lengthen the specification.
   
\subsection{Formal Specifications}
The first formal specification for \amba\ \ahb\ arbiter was given 
in~\cite{DBLP:conf/date/BloemGJPPW07}. 
We have systematically re-written the specifications to make it more 
complete. 
The two important changes are as follows:
(a)~the \htrans [1:0] signal, which plays an important role in \ahb\ transfers, 
was not used in earlier specifications, whereas with the use of \htrans\ signal, 
we make the formal specifications more complete; and  
(b)~the other important change from the specifications of~\cite{DBLP:conf/date/BloemGJPPW07} 
is related to de-assertion of \hlock\ signal: 
according to ARM~\cite{faqarm}, the \ahb\ Master should deassert the \hlock\ 
signal when the address phase of the last transfer in the locked sequence has 
started. 

\begin{figure}[htp]
\includegraphics[width=\linewidth, height=0.5\textheight]{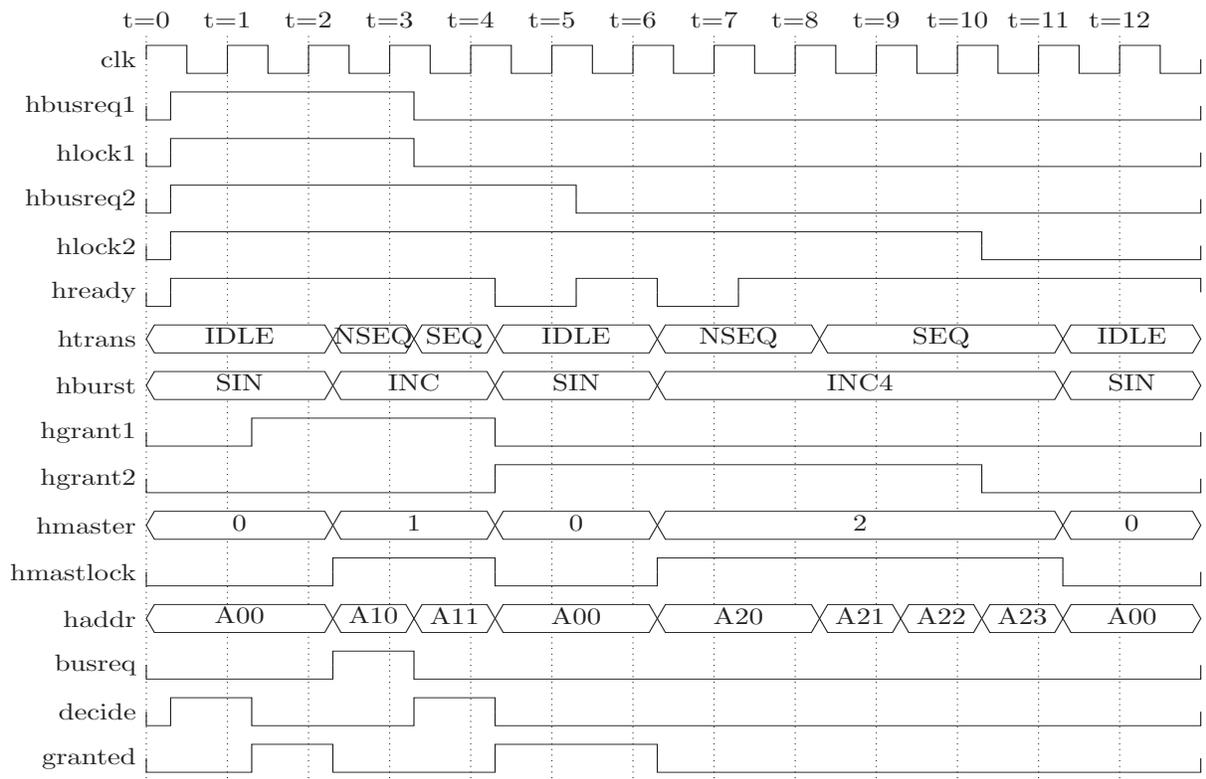}
\caption{\small Signals for the AHB Arbiter and timing diagram}
\end{figure}

Along with the signals described above, we use two auxilary signals 
\decide\ and \busreq, that were introduced in~\cite{DBLP:journals/entcs/BloemGJPPW07}. 
The signal \decide\ indicates the time slot in which the 
arbiter decides who the next master will be and whether its 
access will be locked. The decision is based on \hbusreqi\ and \hlocki. 
The signal \busreq\ points to the \hbusreqi\ signal of the master 
that currently owns the bus.
Two auxilary variables \start\ and \locked, that were introduced 
in~\cite{DBLP:conf/date/BloemGJPPW07}, are not used in our specification. 
It is because with the inclusion of \htrans\ signal and change of nature in
de-assertion of \hlock\ signal, \start\ and \locked\ have become redundant.
We introduce a new auxilary variable \granted\ which is driven by the arbiter. 
The signal \granted\ is used for deciding start of new access. 
When both \granted\ and \hready\ signals are high simultaneously, 
new access shall start in next cycle. Thus a decision can be 
executed at the earliest two time steps after the \hbusreqi\ 
and \hlocki\ signals are read.

We follow the convention used in~\cite{DBLP:conf/date/BloemGJPPW07}: 
guarantees are properties that the arbiter must fulfill, and assumptions are 
properties that the arbiter's environment must fulfill. Our specification for 
the arbiter consists of \emph{9 assumtions} and \emph{12 guarantees} whereas the 
specification from paper~\cite{DBLP:conf/date/BloemGJPPW07} had \emph{4 
assumptions} and \emph{11 guarantees}. 
Figure 2 shows timing diagram for AHB arbiter signals. 
Table 1 contains formal specification of arbiter in \psl. 
The bold faced \textbf{A} and \textbf{G} signify new/re-written property 
whereas non-bold faced indicate existing property 
from~\cite{DBLP:journals/entcs/BloemGJPPW07}. 
The assumptions(A) and guarantees(G) for the arbiter are described below.

\begin{table}[h!]
\caption{\psl\ Specifications for AHB Arbiter.}
\small{
\begin{tabular}{p{.025\textwidth} p{.975\textwidth}}
\toprule

                A1  & $\forall i:$ \always\ ((\hmastlock\ $\wedge$ \hburst\ = \incr) $\to$ ( \nxt\ \eventually\ $\neg$\busreq)) \\ \toprule                  
                A2  & \always\  \eventually\ \hready \\ \toprule 
                A3  & $\forall i:$ \always\ (($\neg$\hbusreqi\ $\wedge$ $\neg$\hlocki\ $\wedge $ (\nxt\ \hlocki))  $\to$ (\nxt\ \hbusreqi))  \\ \toprule 
\multirow{2}{*}{A4} & \always\ ($\neg$\hready\ $\to$ (\htrans\ \eqj\ $\leftrightarrow$ \nxt\ \htrans\ \eqj))\\
                    & \always\ ($\neg$\hready\ $\to$ (\hburst\ \eqj\ $\leftrightarrow$ \nxt\ \hburst\ \eqj))\\ \toprule 
                A5  & \always\ ((\htrans\ = \idle) $\to$ (\nxt\ (\htrans\ $\neq$ \seq)))\\ \toprule 
                A6  & \always\ (((\htrans\ = \nonseq) $\wedge$ (\hburst\ = \incrf) $\wedge$ \hready) $\to$ (\nxt\ (\htrans\ = \seq)))\\ \toprule 
                A7  & \always\ ((\granted\ $\wedge$ \hready) $\to$ (\nxt\  (\htrans\ = \nonseq)))\\ \toprule 
                A8  & \always\ (($\wedge_{\textrm{i=0}}^{\textrm{n-1}} \neg$\hbusreqi) $\to$ (\htrans\ = \idle))\\ \toprule                     
                A9  & $\forall i: (\neg$\hbusreqi\ $\wedge \neg$\hlocki\ $\wedge \neg$\hready\ $\wedge$ (\htrans\ = \idle) $\wedge$ (\hburst\ = \single))\\\toprule  
                G1  & $\forall i:$ \always\ ((\hmaster\ \eqi\ ) $\to$ (\busreq\ $\leftrightarrow$ \hbusreqi))\\ \toprule 
\multirow{2}{*}{G2} & $\forall i:$ \always\ ((\hmastlock\ $\wedge$ (\hburst\ = \incr) $\wedge $ \hready\ $\wedge$ (\htrans\ = \nonseq)) $\to$\\ 
                    & \nxt\ ((\htrans\ = \seq) \untl\ $\neg$\busreq))\\ \toprule 
\multirow{2}{*}{G3} & $\forall i:$ \always\ ((\hmastlock\ $\wedge$ (\hburst\ = \incrf) $\wedge$ \hready\ $\wedge$ (\htrans\ = \nonseq)) $\to $\\
                    & \nxt\ ((\htrans\ = \seq) \untl[3] \hready))\\ \toprule 
                G4  & \always\ ((\decide\ $\wedge (\vee_{\textrm{i=0}}^{\textrm{n-1}}$ \hbusreqi)) $\to$ (\nxt\ \granted ))\\ \toprule 
\multirow{2}{*}{G5} & \always\ ((\granted\ $\wedge\neg$\hready ) $\to$ (\nxt\ \granted)) \\
                    & \always\ ((\granted\ $\wedge$ \hready) $\to$ (\nxt\ $\neg$\granted)) \\ \toprule 
                G6  & $\forall i:$ \always\ (\hready\ $\to$ (\hgranti\ $\leftrightarrow$ \nxt\ (\hmaster\ \eqi)))\\ \toprule 
                G7  & \always\ ((\hready\ $\wedge (\vee_{\textrm{i=0}}^{\textrm{n-1}}$ (\hlocki\ $\wedge$ \hgranti)) $\to$ \nxt\ (\hmastlock)))\\ \toprule 
\multirow{2}{*}{G8} & $\forall i:$ \always\ (($\neg$\hready $\vee$ $\neg$\granted) $\to$ (\hmaster\eqi $\leftrightarrow$ \nxt\ \hmaster\eqi))\\
                    & $\forall i:$ \always\ (($\neg$\hready $\vee$ $\neg$\granted) $\to$ (\hmastlock $\leftrightarrow$ \nxt\ \hmastlock))\\ \toprule 
                G9  & $\forall i:$ \always\ ($\neg$\decide\ $\to$ (\hgranti\ $\leftrightarrow$ \nxt\ \hgranti))\\ \toprule 
\multirow{2}{*}{G10}& $\forall i \neq 0:$ \always\ ($\neg$\hgranti\ $\to$ (\hbusreqi\ \bfore\ \hgranti))\\
                    & \always\ (\decide\ $\wedge \forall i: \neg$\hbusreqi\ $\to$ \nxt\ \hgrantz)\\ \toprule 
                G11 & $\forall i:$ \always\ (\hbusreqi\ $\to$ \eventually\ ($\neg$\hbusreqi\ $\vee$ (\hmaster\ \eqi)))\\ \toprule 
                G12 & \decide\ $\wedge$ \hgrantz\ $\wedge$ (\hmaster\ \eqz) $\wedge$ $\neg$\granted\ $\wedge \neg$\hmastlock\ $\wedge \forall i \neq 0: \neg$\hgranti \\ \bottomrule
\end{tabular}
}
\end{table}

\smallskip\noindent{\bf Assumptions} The assumptions are as follows.
\begin{description*}
\item[{\rm A1}]During a locked unspecified length burst, leaving \hbusreqi\ high locks the bus. This is forbidden by the standard.
\item[{\rm A2}]Leaving \hready\ low locks the bus, the standard forbids it.
\item[{\rm A3}]Signals \hlocki\ and \hbusreqi\ are asserted by \ahb\ Master 
at the same time.
\item[A4]When \hready\ signal is low, all control signals should hold their 
values.
\item[A5]If no transfer is taking place, \htrans\ signal can not become \seq\ 
in the next cycle.
\item[A6]In burst sequence (i.e. \hburst\ = \incrf), if \hready\ is high, 
\nonseq\ transfer shall always be followed by \seq\ transfer.
\item[A7]First transfer of any \ahb\ sequence is \nonseq\ in nature.
\item[A8]When none of the \ahb\ Masters is making a request for bus, 
no transfer will take place.
\item[{\rm A9}]All input signals are low initially.
\end{description*}

\smallskip\noindent{\bf Guarantees} The guarantees are as follows.
\begin{description*}
\item[{\rm G1}]Variable \busreq\ points to \hbusreqi\ of the master that is 
currently granted access to the bus.
\item[G2]When a locked unspecified length burst starts, a new access does not start until the currentmaster (i) releases the bus by lowering HBUSREQi. 
\item[G3]When a length-four locked burst starts, no other accesses start until 
the end of the burst. We can only transfer data when \hready\ is high, so 
the current burst ends at the fourth occurrence of \hready.
\item[G4]Whenever, there is at least one bus request present and signal 
\decide\ is high, \granted\ gets asserted in the next cycle.
\item[G5]If \hready\ is low, then \granted\ signal holds its value. 
Whereas, if \hready\ and \granted\ signals are simultaneously high, then
\granted\ gets deasserted in next cycle.
\item[{\rm G6}]The \hmaster\ signal follows the grants: When \hready\ is high, 
\hmaster\ is set to the master that is currently granted. It implies that no 
two grants may be high simultaneously and the arbiter cannot change 
\hmaster\ without giving a grant.
\item[G7]Whenever signal \hready, \hlocki\ and \hgranti\ are simultaneously 
high, \hmastlock\ gets asserted in the following cycle.
\item[G8]When any of \granted\ or \hready\ signals is low, the 
\hmaster\ and \hmastlock\ signals  do not change. 
\item[{\rm G9}]Whenever \decide\ is low, \hgranti\ signal do not change.
\item[{\rm G10}]We do not grant the bus without a request, except to Master 0. 
If there are no requests, the bus is granted to Master 0.
\item[{\rm G11}]We have a fair bus i.e. every master that has made a 
request shall be serviced eventually.
\item[{\rm G12}]The signals \decide\ and \hgrantz\ are high at first clock 
tick and all others are low.
\end{description*}

Assumptions A1, A2, A3, A9 and Guarantees G1, G2, G3, G6, G8, G9, G10, G11, G12 
mentioned above are taken directly from~\cite{DBLP:conf/date/BloemGJPPW07}. 
Remaining guarantees in~\cite{DBLP:conf/date/BloemGJPPW07} were related to 
auxilary signals which have become redundant in our case with inclusion of \htrans\ 
signal. Out of the above, G2, G3 and G8 have been re-written with the 
original meaning kept intact. Thus all assumptions and guarantees 
from~\cite{DBLP:conf/date/BloemGJPPW07} are taken care in our specification, 
and along with it we have more assumptions and guarantees.

\subsection{Synthesis Results}

\begin{table}[h!]
\begin{tabular}{| p{.1\linewidth} | p{.18\linewidth} | p{.18\linewidth} | p{.16\linewidth} | p{.16\linewidth} |}
\hline
Num of Masters & Synthesis time (sec) from Fig~8 in~\cite{DBLP:journals/entcs/BloemGJPPW07} & 
Synthesis time (sec) for specifications~\cite{DBLP:journals/entcs/BloemGJPPW07} in our experiments & 
Minimum synthesis time (sec) of the last two columns & 
Synthesis time(sec) for our specifications \\\hline
2 & 2 & 2 & 2 & 1 \\\hline
3 & 20 & 22 & 20 & 5 \\\hline
4 & 100 & 103 & 100 & 9 \\\hline
5 & 200 & 203 & 200 & 53 \\\hline
6 & 800 & 677 & 677 & 86\\\hline
7 & 2400 & 2696 & 2400 & 206 \\\hline
8 & 12000 & 7931 & 7931 & 146 \\\hline
9 & 2000 & 2533 & 2000 & 550 \\\hline
10 & 19000 & 18789 & 18789 & 630\\\hline
11 &  &  &  & 577\\\hline
12 &  &  &  & 992\\\hline
13 &  &  &  & 1610\\\hline
14 &  &  &  & 2100\\\hline
15 &  &  &  & 3486\\\hline
16 &  &  &  & 3630\\\hline
\end{tabular}
\caption{Synthesis time comparison}\label{tab2}
\end{table}

Anzu~\cite{DBLP:conf/cav/JobstmannGWB07} is used to synthesize the circuit from
specifications. 
Table~\ref{tab2} shows comparison of time taken by Anzu tool to synthesize
AHB arbiter for different specifications. 
First column shows number of masters for which arbiter was synthesized. 
Second column shows data taken from Figure~8 
of~\cite{DBLP:journals/entcs/BloemGJPPW07} and 
third column shows time taken in synthesizing specification 
from~\cite{DBLP:journals/entcs/BloemGJPPW07} on our machine (2GB RAM). 
In fourth column, we have taken the minimum of these two columns to have the 
best possible estimate of synthesis time for arbiter specifications 
in~\cite{DBLP:journals/entcs/BloemGJPPW07}. 
Fifth column shows the time in seconds for the arbiter synthesized 
using our formal specifications. 

The results (Table~\ref{tab2}) show that using the earlier specifications 
from~\cite{DBLP:journals/entcs/BloemGJPPW07}, the synthesis procedure fails 
for more than 10 masters. 
With our improved specifications we can synthesize arbiter serving upto 16 masters
nearly in an hour.
The \ahb\ standard allows for maximum 16 masters, and arbiter synthesized using our 
specifications can serve upto 16 masters. 
Thus we are able to syntesize arbiter serving the maximum number of masters as 
required by the protocol.
Moreover, our improved specifications show significant (order of magnitude) 
improvement over the earlier specification: for example, for arbiter serving 10 masters 
the synthesis of earlier specifications takes nearly 5 hours, whereas our specification
is synthesized in less than 11 minutes.

\begin{table}[h!]
\begin{tabular}{| p{.1\linewidth} | p{.18\linewidth} | p{.18\linewidth} | p{.16\linewidth} | p{.16\linewidth} |}
\hline
Num of Masters & Gate count from Fig~9 in~\cite{DBLP:journals/entcs/BloemGJPPW07} & 
Gate count for specifications~\cite{DBLP:journals/entcs/BloemGJPPW07} in our experiments & 
Minimum gate count of the last two columns & 
Gate count for our specifications \\\hline
2 & 1000 & 982 & 982 & 182 \\\hline
3 & 3500 & 2626 & 2626 & 409 \\\hline
4 & 8500 & 6801 & 6801 & 776 \\\hline
5 & 11000 & 9033 & 9033 & 920 \\\hline
6 & 18000 & 12448 & 12448 & 1443 \\\hline
7 & 15000 & 19777 & 15000 & 2015 \\\hline
8 & 36000 & NM & 36000 & 2431 \\\hline
9 & NA & 50012 & 50012 & 3047 \\\hline
10 & 50000 & 45912 & 45912 & 2825 \\\hline
11 &  &  &  & 2994\\\hline
12 &  &  &  & 5178\\\hline
13 &  &  &  & 3712\\\hline
14 &  &  &  & 4112\\\hline
15 &  &  &  & 4199\\\hline
16 &  &  &  & 6056\\\hline
\end{tabular}
\caption{Gate count comparison}\label{tab3}
\end{table}

In Table~\ref{tab3}, NA corresponds to not available and NM refers to not mappable by \abc.

Anzu~\cite{DBLP:conf/cav/JobstmannGWB07} tool generates a file in .blif format. 
This file is mapped by using \abc~\cite{abctool} to standard library. \abc\ tool is 
also useful for counting number of gates required to realize the circuit. 
Table~\ref{tab3} shows comparison of number of gates mapped by \abc\ for realizing 
different specifications for arbiter. First column shows the number of masters for 
which the arbiter is synthesized. Second column shows data taken from Figure 8 
in~\cite{DBLP:journals/entcs/BloemGJPPW07} and third column shows number of gates 
mapped by \abc\ tool on our machine (2GB RAM) for existing specification in~\cite{DBLP:journals/entcs/BloemGJPPW07}. 
In fourth column, we have taken the minimum of second and third columns to have 
a best estimate of number of gates for existing specifications. In fifth column, 
gate count for our circuit synthesized from our specification.
Table~\ref{tab3} shows that arbiter synthesized using specifications 
from~\cite{DBLP:journals/entcs/BloemGJPPW07} serving 10 masters has nearly forty-six 
thousand gates, whereas, the \ahb\ arbiter synthesized with our specifications 
serving 10 masters has  only three thousand gates, and even arbiter serving 16 
masters needs only six thousand gates. 
Thus our specifications not only improve the time taken for synthesizing, but 
also improve the gate count of synthesized circuit by an order of magnitude.

\begin{figure}[h!]
\begin{minipage}[b]{0.5\linewidth} 
\centering
\includegraphics[width=9cm]{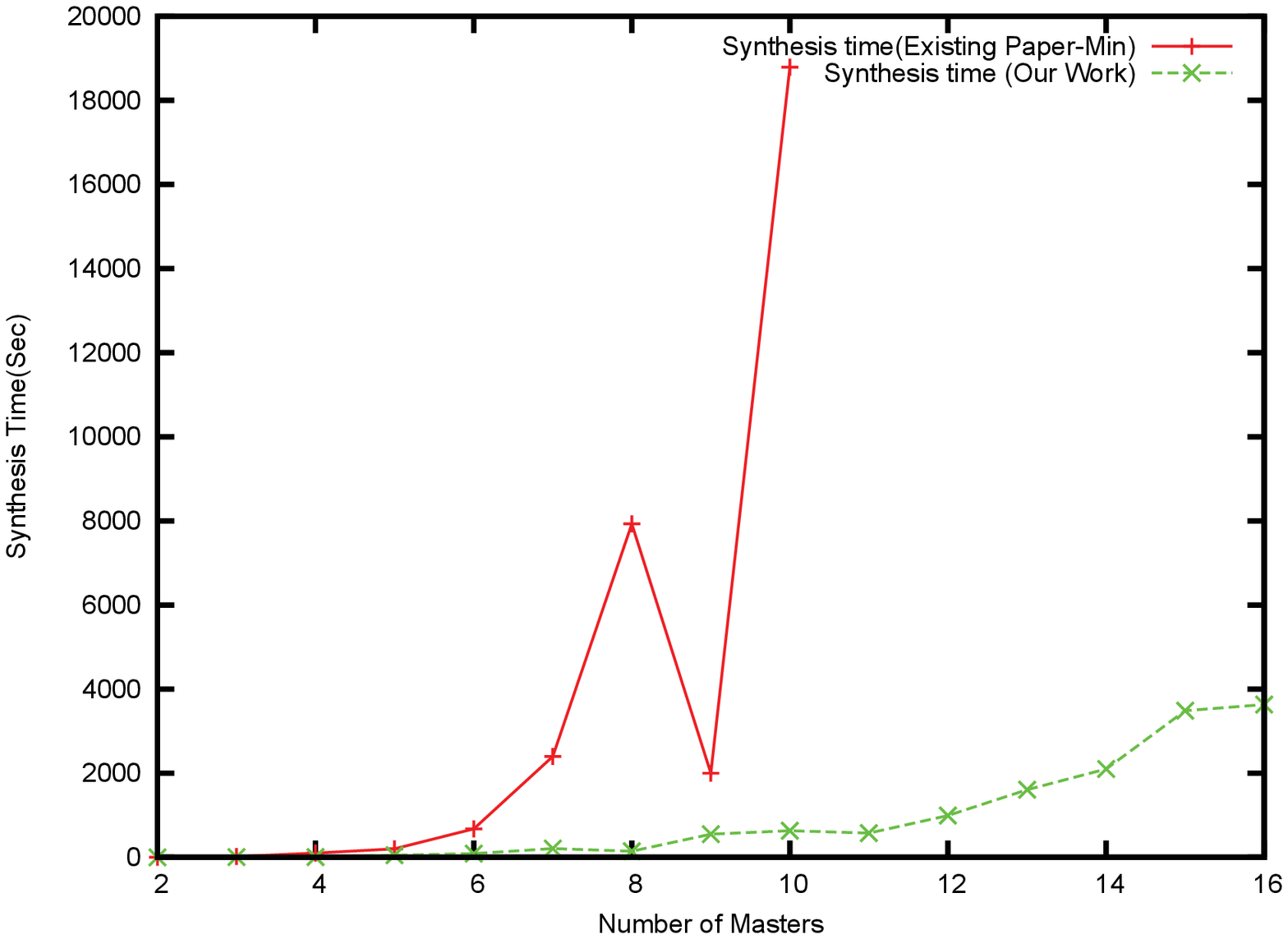}
\caption{\small Synthesis Time Comparison.}\label{fig_3}
\end{minipage}
\hspace{0.5cm} 
\begin{minipage}[b]{0.5\linewidth}
\centering
\includegraphics[width=9cm]{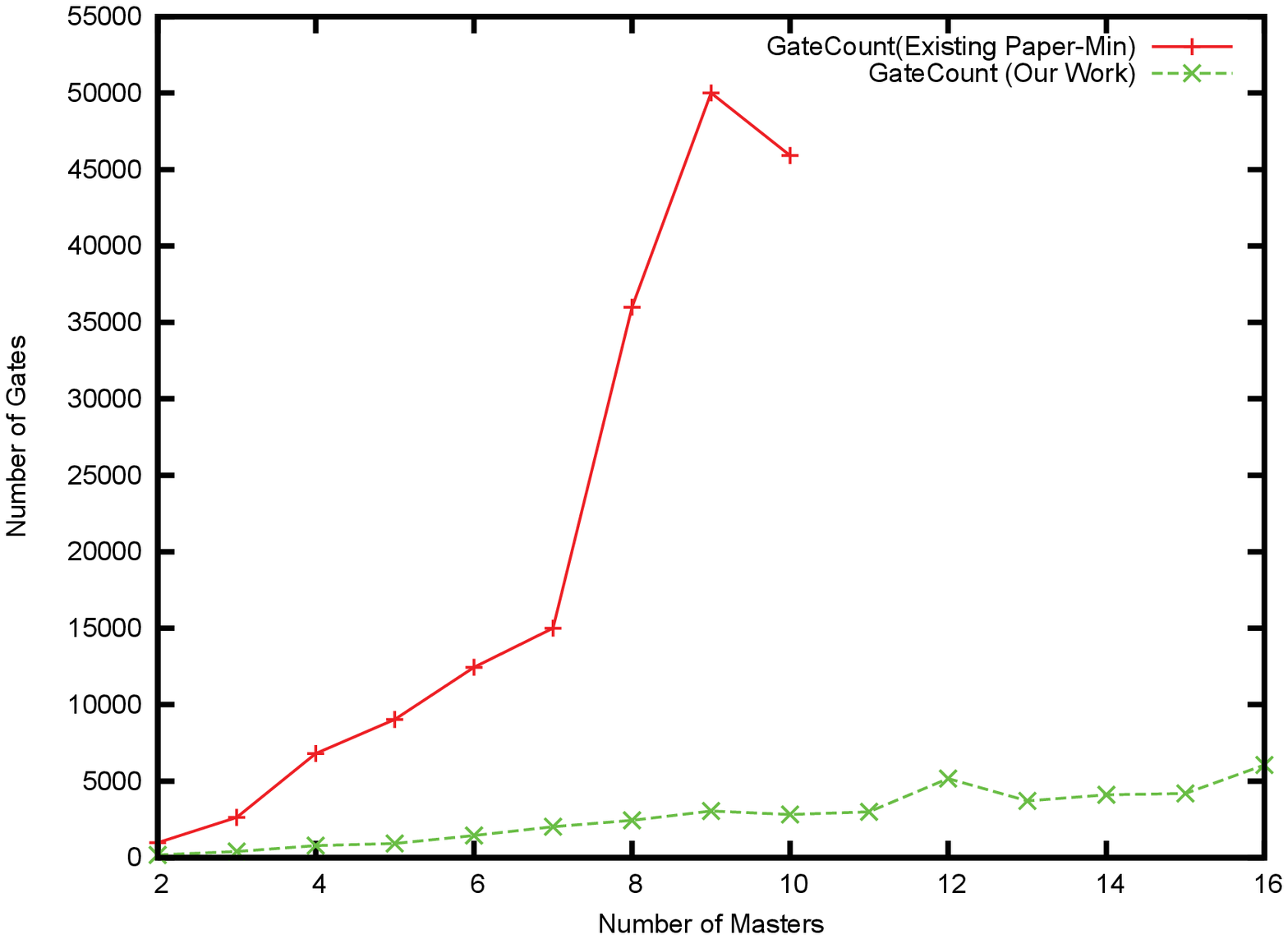}
\caption{\small Gate Comparison.}\label{fig_4}
\end{minipage}
\end{figure}

Graphical comparison for arbiters serving different number of masters is shown in 
Figure~\ref{fig_3} and Figure~\ref{fig_4}. 
Figure~\ref{fig_3} shows comparison for synthesis time whereas 
Figure~\ref{fig_4} depicts comparison for gate count.

\section{AMBA AHB Master} 
In this section we present the synthesis results for AHB Master: we first 
present the signals, then the specification, and then the synthesis results.

\subsection{AHB Master Signals}
We first introduce the signals for \ahb\ Master that have not been introduced.
\begin{itemize*}
\item \hwrite\ - This signal from bus master indicates nature of transfer. When \hwrite\ is low, it indicates read transfer. If high, it indicates write transfer.
\item \haddr[31:0] - These signals from the master provide information about location where write or read transfer shall take place.
\item \hwdata[31:0] - These signals from the master provide information about data to be written in case of write transaction.
\item \hrdata[31:0] - These signals from bus slave to bus master provide information about data read in case of read transaction.
\item \hsze[2:0] - This signal from bus master to bus slave provides information about bus width. It can be one of byte(8-bit), half-word(16-bit), word(32-bit) and up to 1024 bits. We have fixed it to word i.e. data bus shall be 32-bit wide. 
\item \hresp[1:0] - This signal from bus slave to bus master provides transfer response. It can be one of \okay, \error, \splt\ and \retry. \splt\ and \retry\ are optional feature allowed in standard. For simplicity, we have fixed it to \okay\ otherwise it would lengthen the specification.
\end{itemize*}

\begin{figure}[h!]
\includegraphics[width=\linewidth, height=0.45\textheight]{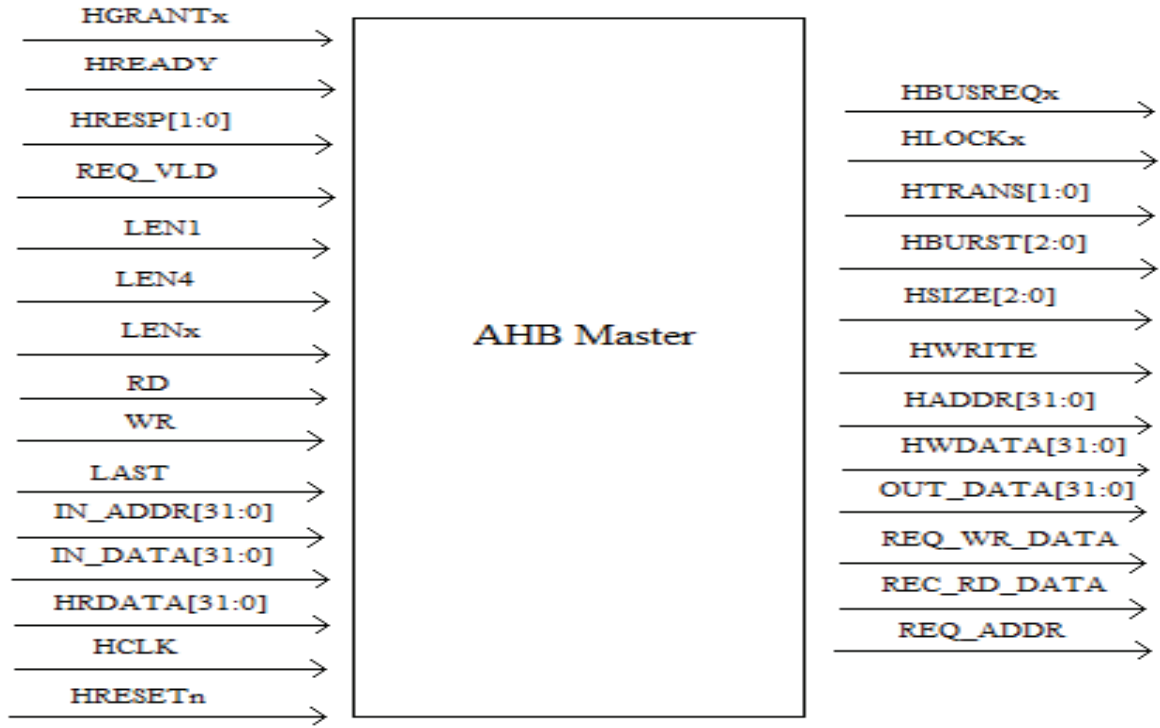}
\caption{\small AHB Master}\label{fig_5}
\end{figure}
The \amba\ \ahb\ specification also allows protection controls but for simplicity, 
we have left that feature out. 
Few auxilary signals are also used. They are as follows:
\begin{itemize*}
\item \reqvld\ - This signal is input to bus master. 
It is used by bus master for deciding \hbusreq. \hbusreq\ 
signal is asserted whenever \reqvld\ is asserted.
\item \wrt\ - This signal is input to bus master. 
It indicates that write transaction shall take place. \hwrite\ shall be 
\high\, if  \wrt\ is high.
\item \re\ - This signal is input to bus master. 
If high, it indicates that read transaction shall take place and hence \hwrite\ 
shall be set \low.
\item \leno\ - This signal is input to bus master. It indicates that single transfer shall take place.
\item \lenf\ - This signal is input to bus master. It informs that the transfer should be a burst sequence of four transfers.
\item \lenu\ - This signal is input to bus master. It informs that the transfer should be a burst sequence of unspecified length.
\item \inaddr[31:0] - These signals are input to the master providing information about address. These signals are used to decide \haddr. 
\item \indata[31:0] - These signals are input to the master providing information about write data. These signals are used to decide \hwdata.
\item \last\ - This signal is input to bus master. It indicates the last transfer in a sequence of transfers.
\item \outdata[31:0] - These signals from the master provide information about read data.
\item \reqaddr\ - This signal from the master is request for address. 
If this signal is high, in the next clock cycle, master shall receive \inaddr.
\item \reqwrdata\ - This signal from the master is request for data. 
If this signal is high, in the next clock cycle, master shall receive \indata.
\item \recrddata\ - This signal from the master provides acts as valid signal for read data.
If it is high, \hrdata\ shall be copied to \outdata.
\end{itemize*}

\begin{figure}[h!]
\includegraphics[width=\linewidth, height=0.7\textheight]{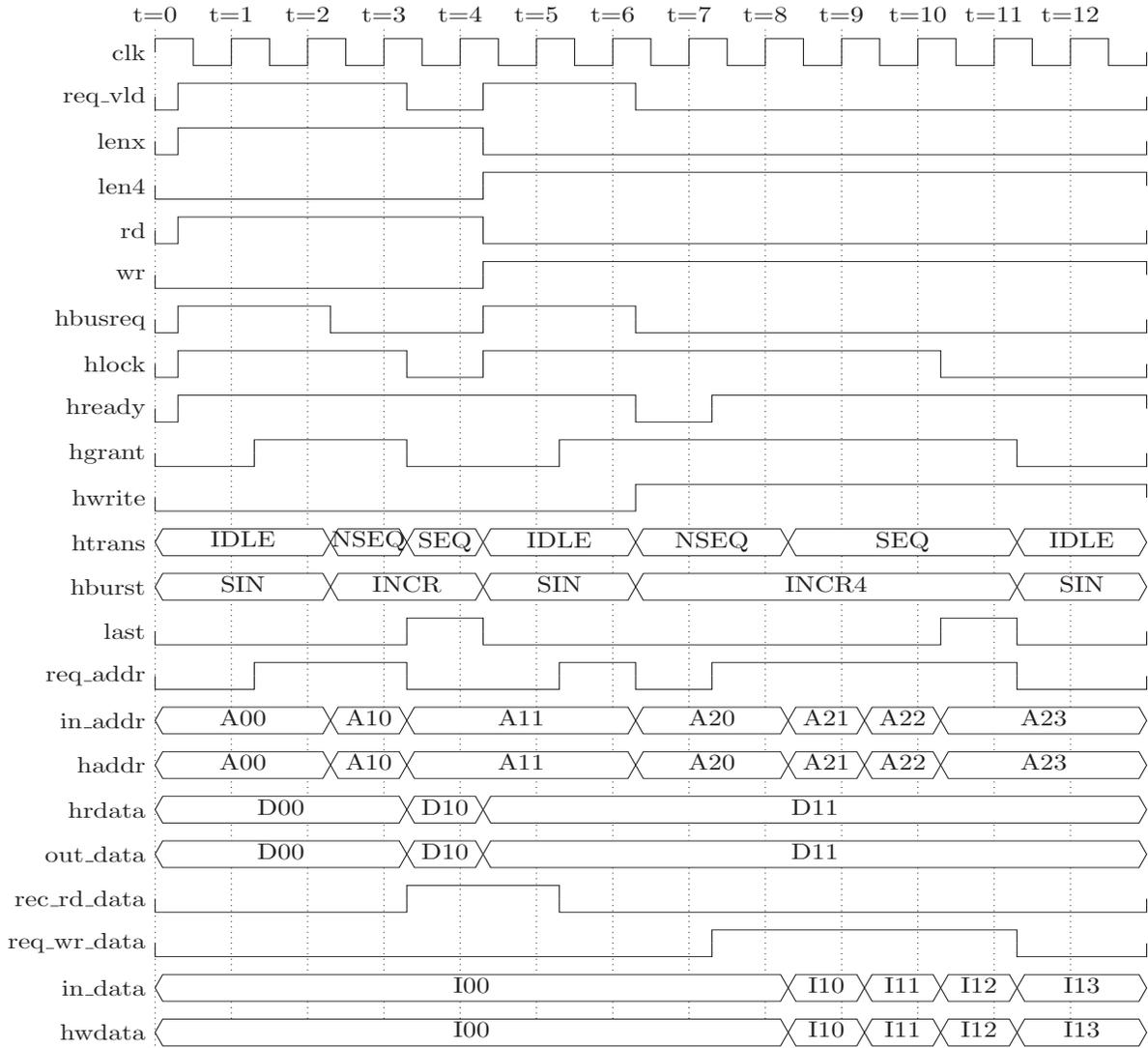}
\caption{\small Signals for the \ahb\ Master}\label{fig_6}
\end{figure}
Figure~\ref{fig_5} shows signals for AHB Master and Figure~\ref{fig_6} shows timing diagram for those signals.

\subsection{Formal Specifications}
In the formal specification of \amba\ \ahb\ Master, we have 10 assumptions and 
15 guarantees. 

\smallskip\noindent{\bf Assumptions} The assumptions are as follows.
\begin{description*}
\item[A1]Length of transfer will be specified with \reqvld\ signal i.e. 
whenever \reqvld\ is high, one of \leno, \lenf\ and \lenu\ signal shall be high.
\item[A2]Nature of transfer will be specified with \reqvld\ signal i.e. 
whenever \reqvld\ signal is high, one of \re\ and \wrt\ signal shall be high. 
\item[A3]If \reqvld signal is low, \re, \wrt, \leno, \lenf\ and \lenu\ 
shall hold their values.
\item[A4]There can not be conflict between signals indicating nature of 
transfer thus \re\ and \wrt\ signal can not be high simultaneously.
\item[A5]There can not be conflict between signals indicating length of 
transfer thus \leno, \lenf\ and \lenu\ signals can not be high simultaneously.
\item[A6]Input \hresp\ signal shall be \okay\ throughout.
\item[A7]The bus is fair one, hence every \hbusreq\ shall eventually be 
answered. 
\item[A8]During a locked unspecified length burst, leaving \hbusreq\ 
high locks the bus. This is forbidden by the standard.
\item[A9]Eventually \hready\ will be high.
\item[A10]We are not considering it as default bus master for the sake of 
generality. Hence eventually  \reqvld\ and \hgrant\ signals will be low.
\end{description*}

We are assuming that data bus is 32-bit wide, hence \hsze\ will be fixed to 
\word. 
To make this bus master more general, another assumption is that this 
bus master requests for only locked transfers.

\smallskip\noindent{\bf Guarantees} The guarantees are as follows.
\begin{description*}
\item[G1]Data bus is 32-bit wide. Thus \hsze\ shall be fixed to \word\ 
throughout.
\item[G2]\hbusreq\ signal gets asserted and de-asserted with \reqvld.
\item[G3]Bus master requests only for locked transfer.
\item[G4]If the ongoing transfer is last transfer of an ahb sequence, 
\hlock\ shall be lowered.
\item[G5]Length four burst (\hburst\ = \incrf) shall end at fourth occurence 
of \hready.
\item[G6]\hburst\ shall be set according to length of the transfer indicated by \leno, \lenf\ and \lenu.
\item[G7]First transfer of an \ahb\ sequence is always \nonseq\ in nature. 
All following transfers in sequence shall be \seq\ in nature.
\item[G8]Nature of transfer shall be set according to \wrt\ and \re\ signals. 
\item[G9]If \hready\ is low, all control signals shall hold their values.
\item[G10]When \hready\ and \hgrant\ are simultaneously high, \reqaddr\ signal 
shall be high. 
It ensures that in next cycle, master can put address on address bus.
\item[G11]When both \reqaddr\ and \wrt\ signals are high, \reqwrdata\ 
signal shall also be high. It ensures that data shall be put on data 
bus one cycle after address is put on address bus.
\item[G12]When a read transfer is taking place and \hready\ is high, 
\recrddata\ signal shall also be high.
\item[G13]When \reqaddr\ is high, in the next cycle, incoming \inaddr\ 
shall be copied to address bus.
\item[G14]When \reqwrdata\ is high, in the next cycle, incoming \indata\ 
shall be copied to data bus.
\item[G15]When read transaction is in progress and \hready\ is high, \outdata\ 
shall copy the value of \hrdata.
\end{description*}

\begin{table}[htp]
\caption{\psl\ Specifications for AHB Master}
\small{
\begin{tabular}{p{.025\textwidth} p{.975\textwidth}}
\toprule
                A1  & \always\ (\reqvld\ $\to$ (\lenu\ $\vee$ \leno\ $\vee$ \lenf))\\ \toprule                   
                A2  & \always\ (\reqvld\ $\to$ (\wrt\ $\vee$ \re))\\ \toprule     
\multirow{5}{*}{A3} & \always\ ((\nxt\ $\neg$\reqvld) $\to(\neg$\leno\ $\leftrightarrow$ (\nxt\ $\neg$\leno)))\\
                    & \always\ ((\nxt\ $\neg$\reqvld) $\to(\neg$\lenu\ $\leftrightarrow$ (\nxt\ $\neg$\lenu)))\\
                    & \always\ ((\nxt\ $\neg$\reqvld) $\to(\neg$\lenf\ $\leftrightarrow$ (\nxt\ $\neg$\lenf)))\\
                    & \always\ ((\nxt\ $\neg$\reqvld) $\to(\neg$\wrt\  $\leftrightarrow$ (\nxt\ $\neg$\wrt)))\\
                    & \always\ ((\nxt\ $\neg$\reqvld) $\to(\neg$\re\   $\leftrightarrow$ (\nxt\ $\neg$\re)))\\ \toprule     
\multirow{2}{*}{A4} & \always\ (\wrt\ $\to \neg$ \re)\\                                    
                    & \always\ (\re\ $\to \neg$ \wrt)\\ \toprule                                   
\multirow{3}{*}{A5} & \always\ (\lenu\ $\to (\neg$\leno\ $\vee$ $\neg$\lenf))\\ 
                    & \always\ (\leno\ $\to (\neg$\lenu\ $\vee$ $\neg$\lenf))\\
                    & \always\ (\lenf\ $\to (\neg$\lenu\ $\vee$ $\neg$\leno))\\ \toprule                   
                A6  & \always\ (\hresp\ = \okay)\\ \toprule            
                A7  & \always\ (\reqvld\ $\to$ \eventually\ \hgrant)\\ \toprule 
                A8  & \always\ ((\hlock\ $\wedge$ (\hburst\ = \incr)) $\to$ \nxt\ \eventually\ $\neg$\reqvld)\\ \toprule 
                A9  & \always\ (\eventually\ \hready)\\ \toprule            
                A10 & \always\ (\eventually\ ($\neg$\reqvld\ $\wedge$ $\neg$\hgrant))\\ \toprule  
                G1  & \always\ (\hsze\ = \word)\\ \toprule
                G2  & \always\ (\reqvld\ $\to$ \hbusreq)\\ \toprule
                G3  & \always\ (($\neg$\hbusreq\ $\wedge$ \nxt\ \hbusreq\ $\wedge$ $\neg$\hlock) $\to$ \nxt\ \hlock)\\ \toprule
                G4  & \always\ (\last\ $\to \neg$\hlock )\\ 
\multirow{2}{*}{G5} & \always\ ((\hlock\ $\wedge$ (\hburst\ = \incrf) $\wedge$ \hready\ $\wedge$ (\htrans\ = \nonseq)) $\to$\\
                    & \nxt\ ((\htrans\ = \seq) \untl[3] \hready))\\ \toprule 
\multirow{3}{*}{G6} & \always\ (\hbusreq\ $\wedge$ \hgrant\ $\wedge$ (\htrans\ = \idle) $\wedge$ \hready\ $\wedge$ \leno\ $\to$ \nxt\ (\hburst\ = \single))\\
                    & \always\ (\hbusreq\ $\wedge$ \hgrant\ $\wedge$ (\htrans\ = \idle) $\wedge$ \hready\ $\wedge$ \lenu\ $\to$ \nxt\ (\hburst\ = \incr))\\
                    & \always\ (\hbusreq\ $\wedge$ \hgrant\ $\wedge$ (\htrans\ = \idle) $\wedge$ \hready\ $\wedge$ \lenf\ $\to$ \nxt\ (\hburst\ = \incrf))\\ \toprule 
\multirow{3}{*}{G7} & \always\ (\hbusreq\ $\wedge$ \hgrant\ $\wedge$ (\htrans\ = \idle) $\wedge$ \hready\ $\to$ \nxt\ (\htrans\ = \nonseq))\\
                    & \always\ ($\neg$\last\ $\wedge$ (\htrans\ = \nonseq) $\wedge$ \hready\ $\to $ \nxt\ (\htrans\ = \seq))\\
                    & \always\ ((\htrans\ = \idle) $\to$ (\hburst\ = \single))\\ \toprule
\multirow{2}{*}{G8} & \always\ (\hgrant\ $\wedge$ (\htrans\ = \nonseq) $\wedge$ \hready\ $\wedge$ \wrt\ $\to$ \hwrite)\\
                    & \always\ (\hgrant\ $\wedge$ (\htrans\ = \nonseq) $\wedge$ \hready\ $\wedge$ \re\ $\to \neg$\hwrite)\\ \toprule                   
\multirow{2}{*}{G9} & \always\ ($\neg$\hready\ $\to$ ((\htrans\ \eqj) $\leftrightarrow$ \nxt\ (\htrans\ \eqj)))\\
                    & \always\ ($\neg$\hready\ $\to$ ((\hburst\ \eqj) $\leftrightarrow$ \nxt\ (\hburst\ \eqj)))\\ \toprule                   
                G10 & \always\ ((\hready\ $\wedge$ \hgrant) $\to$ \reqaddr)\\ \toprule                    
                G11 & \always\ ((\reqaddr\ $\wedge$ \hwrite) $\to$ \reqwrdata)\\ \toprule                   
                G12 & \always\ ((\hready\ $\wedge$ ((\htrans\ = \nonseq) $\vee $ (\htrans\ = \seq)) $\wedge \neg$\hwrite) $\to$ \recrddata)\\ \toprule       
                G13 & $\forall i:$ \always\ (\reqaddr\ $\to$ ((\nxt\ (\inaddri\ \eqj)) $\leftrightarrow$ (\nxt\ (\haddri\ \eqj))))\\ \toprule                
                G14 & $\forall i:$ \always\ (\reqwrdata\ $\to$ ((\nxt\ (\indatai\ \eqj)) $\leftrightarrow$ (\nxt\ (\hwdatai\ \eqj))))\\ \toprule 
                G15 & $\forall i:$ \always\ (\hready\ $\wedge \neg$\hwrite\ $\wedge$ ((\htrans\ = \seq)\\
                    & $\vee$ (\htrans\ = \nonseq)) $\to$ ((\nxt\ (\hrdatai\ \eqj)) $\leftrightarrow$ (\nxt\ (\outdatai\ \eqj))))\\ \bottomrule
\end{tabular}
}
\end{table}

\subsection{Synthesis Results}
The synthesis time for the \ahb\ Master is 8.3 seconds. The generated circuit 
is mapped using \abc\ tool. It has 157 gates with area 210 square units. 
It is a very small circuit even with respect to manual implementations. 
Thus we are not only able to synthesize the \ahb\ Master from its formal 
specifications, but the synthesized circuit is also very compact.

\section{\amba\ \ahb\ Slave}
In this section we present the synthesis results for AHB Slave.

\subsection{\ahb\ Slave Signals}

\begin{figure}[h]
\includegraphics[width=\linewidth, height=0.4\textheight]{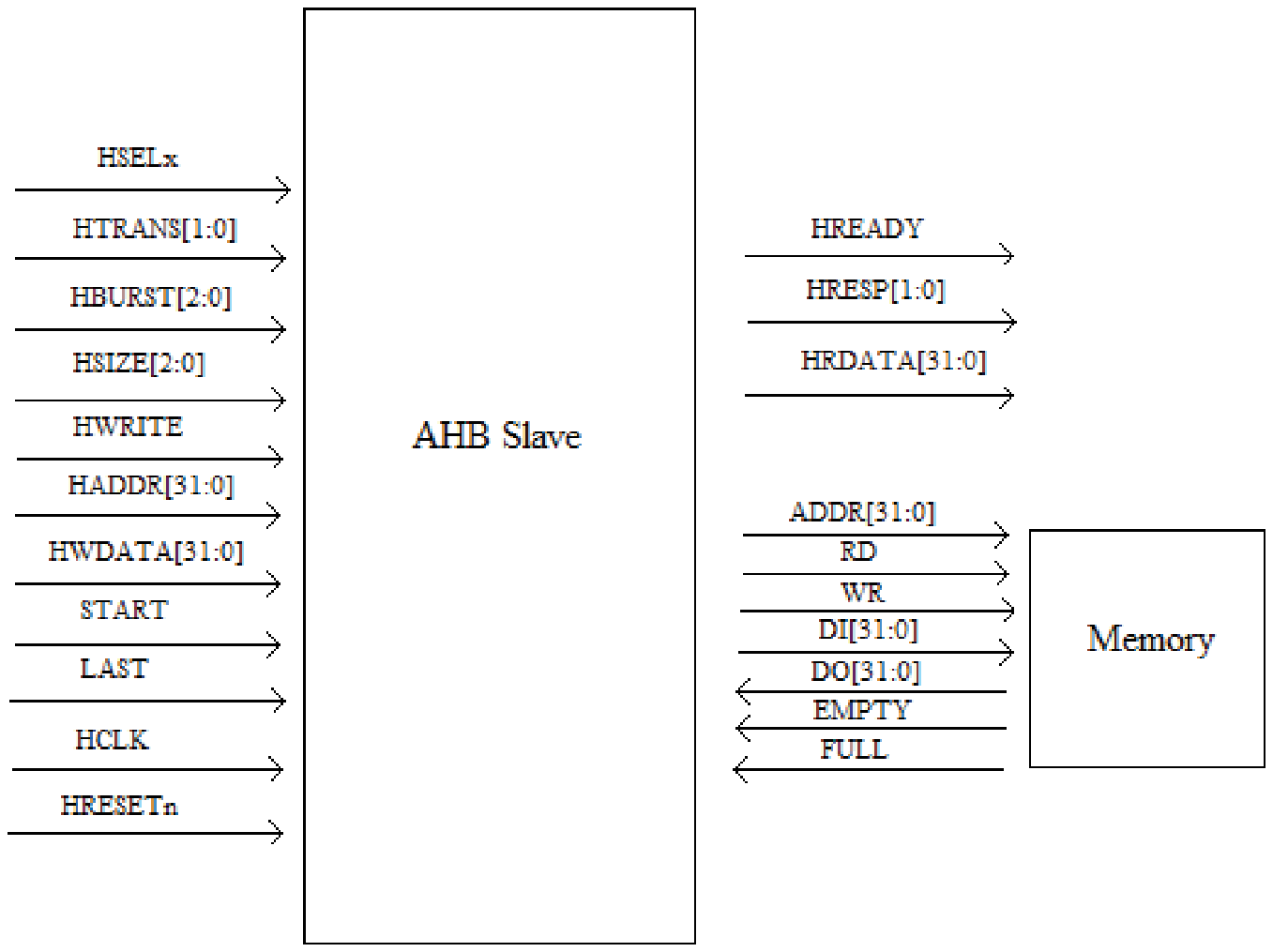}
\caption{\small \ahb Slave}\label{fig7}
\end{figure}
The signals that are useful for \ahb\ slave are already described in previous sections. 
We have introduced an interface between slave and a memory so that read and 
write transactions can be implemented. We are considering memory with two 
status signals \emty\ and \ful. 

Two auxilary signals have also been added named \start\ and \last. \start\ signal 
indicates start of an \ahb\ transfer or sequence whereas \last\ signal is used to 
indicate last transfer of an \ahb\ sequence.

The signals used in this interface are shown in Figure~\ref{fig7}. 
Figure~\ref{fig8} shows the timing diagram from \ahb\ slave signals. The description 
of signals used in interface between slave and memory is given below:

\begin{figure}[h!]
\includegraphics[width=\linewidth, height=0.475\textheight]{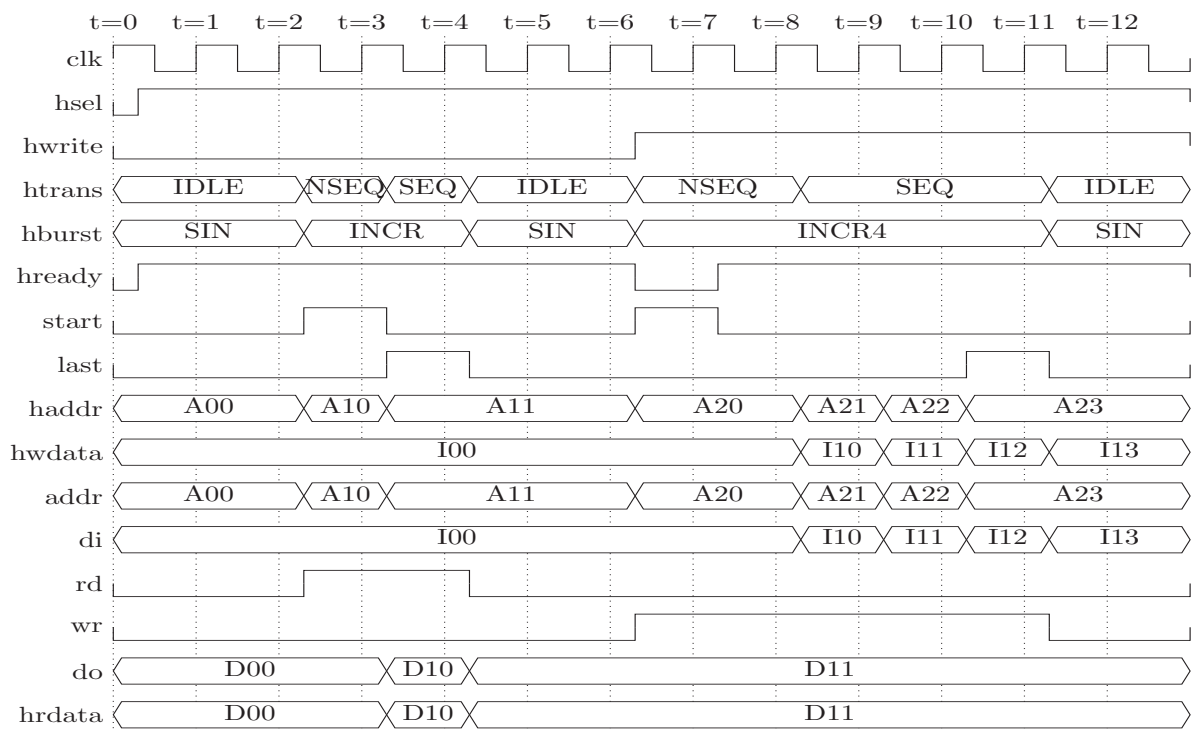}
\caption{\small Signals for the \ahb\ Slave}\label{fig8} 
\end{figure}

\begin{itemize*}
\item \ful\ - This signal is input to bus slave indicating memory is full. No more data can be written into it without first being read. 
\item \emty\ - This signal is input to bus slave indicating memory is empty. No more data can be read from it without first being written. 
\item \addr[31:0] - These signals are output from slave providing address information.
\item \di[31:0] - These signals are output from slave and input to memory providing information about data that should be written into memory.
\item \dout[31:0] - These signals are output from memory and input to slave providing information about data that has been read from memory.
\item \re\ - This signal is input to memory from slave. It indicates that the read operation is being executed.
\item \wrt\ - This signal is input to memory from slave. It indicates that the write operation is being executed.
\end{itemize*} 

\subsection{Formal Specifications}
In the formal specification of \amba\ \ahb\ Slave, we have 7 assumptions and 
9 guarantees. They are as follows.

\begin{table}[h!]
\caption{\psl\ Specifications for \ahb\ Slave}
\small{
\begin{tabular}{p{.01\textwidth} p{.99\textwidth}}
\toprule

                A1  & \always\ ($\neg$\hsel\ $\to$ ((\htrans\ = \idle) $\wedge$ (\hburst\ = \single) $\wedge$ $\neg$\hwrite\ $\wedge$ $\neg$\strt\ $\wedge$ $\neg$\last))\\ \toprule       
                A2  & \always\ ((\htrans\ = \idle) $\to$ ((\hburst\ = \single) $\wedge$ $\neg$\hwrite\ $\wedge$ $\neg$\strt\ $\wedge$ $\neg$\last))\\ \toprule    
                A3  & \always\ (\strt\ $\to$ (\htrans\ = \nonseq))\\ \toprule
                A4  & \always\ ($\neg$\last\ $\wedge$ (\htrans\ = \nonseq) $\wedge$ \hready\ $\to$ \nxt\ (\htrans\ = \seq))\\ \toprule                                       A5  & \always\ ((\hlock\ $\wedge$ (\hburst\ = \incrf) $\wedge$ \hready\ $\wedge$ (\htrans\ = \nonseq)) $\to $ \nxt ((\htrans\ = \seq) \untl[3] \hready))\\ \toprule 
                A6  & \always\ ((\last\ $\wedge$ \nxt\ $\neg$\strt) $\to$ \nxt\ (\htrans\ = \idle))\\ \toprule            
\multirow{5}{*}{A7} & \always\ ($\neg$\hready\ $\to$((\htrans\ \eqj) $\leftrightarrow$ \nxt\ (\htrans\ \eqj)))\\
                    & \always\ ($\neg$\hready\ $\to$((\hburst\ \eqj) $\leftrightarrow$ \nxt\ (\hburst\ \eqj)))\\ 
                    & \always\ ($\neg$\hready\ $\to$((\haddr\  \eqj) $\leftrightarrow$ \nxt\ (\haddr\ \eqj)))\\ 
                    & \always\ ($\neg$\hready\ $\to$((\hwdata\ \eqj) $\leftrightarrow$ \nxt\ (\hwdata\ \eqj)))\\ 
                    & \always\ ($\neg$\hready\ $\to$((\dout\ \eqj)  $\leftrightarrow$ \nxt\ (\dout\ \eqj)))\\ \toprule
                G1  & \always\ ($\neg$\hsel\ $\to$ \hready)\\ \toprule                                
                G2  & \always\ ($\neg$\hsel\ $\to$ (\hresp\ = \okay))\\ \toprule       
                G3  & \always\ ((\htrans\ = \idle) $\to$ (\hresp\ = \okay))\\ \toprule       
\multirow{2}{*}{G4} & \always\ ((\wrt\ $\wedge$ \hsel) $\to \neg$\re)\\
                    & \always\ ((\re\ $\wedge$ \hsel) $\to \neg$\wrt) \\\toprule       
\multirow{2}{*}{G5} & \always\ ((\hsel\ $\wedge$ \ful\ $\wedge$ \wrt) $\to$ (\hresp\ = \error))\\
                    & \always\ ((\hsel\ $\wedge$ \emty\ $\wedge$ \re) $\to$ (\hresp\ = \error))\\ \toprule       
\multirow{2}{*}{G6}& \always\ ((\hsel\ $\wedge$ ((\htrans\ = \nonseq) $\vee$ (\htrans\ = \seq)) $\wedge$ \hwrite) $\to$ \wrt)\\
                   & \always\ ((\hsel\ $\wedge$ ((\htrans\ = \nonseq) $\vee$ (\htrans\ = \seq)) $\wedge \neg$\hwrite) $\to$ \re)\\ \toprule 
                G7 & \always\ ((\hsel\ $\wedge$ ((\htrans\ = \nonseq) $\vee$ (\htrans\ = \seq)) $\to$ ((\haddr\ \eqj) $\leftrightarrow$ (\addr\ \eqj)))\\ \toprule
                G8 & \always\ ((\hsel\ $\wedge$ ((\htrans\ = \nonseq) $\vee$ (\htrans\ = \seq)) $\wedge$ \hwrite) $\to$ ((\hwdata\ \eqj) $\leftrightarrow$ (\di\ \eqj)))\\ \toprule
                G9 & \always\ ((\hsel\ $\wedge$ ((\htrans\ = \nonseq) $\vee$ (\htrans\ = \seq)) $\wedge \neg$\hwrite) $\to$ ((\dout\ \eqj) $\leftrightarrow$ (\hrdata\ \eqj)))\\ \bottomrule
\end{tabular}
}
\end{table}

\smallskip\noindent{\bf Assumptions} The assumptions are as follows.
\begin{description*}
\item[A1]When the slave is not selected by the decoder, all control signals shall be low.
\item[A2]When \htrans\ is \idle, all control signals shall be low.
\item[A3]First transfer of any sequence is \nonseq\ in nature. 
\item[A4]Non-first transfer of an \ahb\ sequence will always be \seq\ in nature.
\item[A5]Burst sequence of length four shall end at fourth occurence of \hready.
\item[A6]If this is last transaction of a sequence and next cycle is not start 
of another sequence, \htrans\ shall be \idle\ in next cycle.
\item[A7]If \hready\ is low, all control signals, address and data buses shall hold their values.
\end{description*}

\smallskip\noindent{\bf Guarantees} The guarantees are as follows.
\begin{description*}
\item[G1]When the slave is not selected by the decoder, \hready\ signal shall be high.
\item[G2]When the slave is not selected by the decoder, \hresp\ shall be \okay.
\item[G3]When no \ahb\ transaction is taking place, \hresp\ shall be \okay.
\item[G4]\re\ and \wrt\ signal can not be high simultaneously.
\item[G5]If memory is full and write tranfer is attempted, slave shall send 
\error\ response. 
Similarly, if memory is empty and read transfer is attempted, slave shall 
send \error\ response.
\item[G6]When slave is involved in a transfer, \hwrite\ is used to decide 
values of \wrt\ and \re.
\item[G7]When slave is involved in any transfer, signal \haddr\ is used to decide \addr.
\item[G8]When slave is involved in write transfer, signal \hwdata\ is used to decide \di.
\item[G9]When slave is involved in read transfer, signal \dout\ is used to decide \hrdata.
\end{description*} 

\subsection{Synthesis Results}
The synthesis time for the AHB Slave is 21.5 seconds. The circuit generated, when mapped using ABC, 
has 214 gates with area 429 unit squared. It is a very small circuit even with respect to 
manual implementations. 
Thus we are not only able to synthesize the AHB Slave from its formal specifications, 
but the synthesized circuit is also very compact.

\section{Lessons Learned}
In the process of systematically re-writing the formal specifications for efficient 
synthesis, we learnt a few lessons about writing formal specifications for synthesis.
We present these lessons with examples below.
\begin{itemize*}
\item In the process of writing specifications, we should first simplify the design 
(if possible), write realizable specification for that can be synthesized efficiently 
for the simple design, and finally add necessary complexities to have the complete 
specification. 
For example, while writing \ahb\ Master specifications, we first fixed all data and address 
signals width to one bit, synthesized the simpler design successfully and efficiently. 
This was followed by increasing data and address signal widths to 32-bit and adding 
necessary changes to \ahb\ Master specifications to make it complete and synthesizable.

\item While writing specifications, proceeding with the execution order of events is helpful. 
For example, while writing \ahb\ Arbiter specifications, we proceeded with writing properties 
related to requesting access, granting access followed by properties related to \ahb\ transfers.

\item The use of auxilary signals is helpful in scenarios that cannot be emulated 
using only input output signals. For example, in \ahb\ Slave specifications, we 
have introduced auxilary signals to emulate slave-memory interactions.

\item The eventual specifications were the most time-consuming and difficult ones for synthesis
and they need special attention. 

\end{itemize*}
In general, most data intensive applications are not reacive designs of degree one, 
and the above approach may not be ideal for those applications,  but we believe that 
the above synthesis approach should work well for many control specific applications.

\paragraph{{\bf Acknowledgment.}} We thank Barbara Jobstmann for 
explaining the changes made in the specifications 
from~\cite{DBLP:conf/date/BloemGJPPW07} 
to~\cite{DBLP:journals/entcs/BloemGJPPW07}.

{\small
\bibliography{main}
\bibliographystyle{plain}
}

\end{document}